\documentclass{aa}
\usepackage{graphics}
\def\a{{$\alpha$}}

\newcommand{\h}{$^{\rm h}$}
\newcommand{\m}{$^{\rm m}$}
\newcommand{\s}{$^{\rm s}$}
\newcommand{\dd}{$\delta$}
\newcommand{\ha}{\rm H$\alpha$}
\newcommand{\hbeta}{\rm H$\beta$}
\newcommand{\HII}{\ion{H}{ii}}
\newcommand{\hnii}{{\rm H}$\alpha+[$\ion{N}{ii}$]$}
\newcommand{\nii}{$[$\ion{N}{ii}$]$}
\newcommand{\sii}{$[$\ion{S}{ii}$]$}

\newcommand{\oiii}{$[$\ion{O}{iii}$]$}

\newcommand{\et}{et al.}
\newcommand{\flux}{$10^{-17}$ erg s$^{-1}$ cm$^{-2}$ arcsec$^{-2}$}
\newcommand{\dens}{\rm cm$^{-3}$}
\newcommand{\vel}{\rm km s$^{-1}$}
\newcommand{\sulfur}{[S~{\sc ii}]}

\newcommand{\oxygen}{[O~{\sc iii}]}
\newcommand{\siirat}{$[$\ion{S}{ii}$]\lambda\lambda\ 6716/6731$} 
\begin{document}

%
\title{Deep optical observations of the supernova remnants\\ G 126.2$+$1.6, G 59.8$+$1.2 and G 54.4$-$0.3}

\author{P. Boumis\inst{1}
\and  F. Mavromatakis\inst{2}
\and E. M. Xilouris\inst{1}
\and J. Alikakos\inst{1,3}
\and M. P. Redman\inst{4}
\and C. D. Goudis\inst{1,3}
}
\offprints{P. Boumis,~~\email{ptb@astro.noa.gr}}
\authorrunning{P. Boumis et al.}
\titlerunning{Deep optical observations of the SNRs G 54.4$-$0.3, G 59.8$+$1.2, G 126.2$+$1.6}
\institute{
Institute of Astronomy \& Astrophysics, National Observatory of
Athens, I. Metaxa \& V. Paulou, P. Penteli, GR-15236 Athens, Greece.
\and University of Crete, Physics Department, P.O. Box 2208, 710 03 Heraklion, Crete, Greece.
\and Astronomical Laboratory, Department of Physics, University of Patras, 26500 Rio-Patras, Greece.
\and Department of Physics, National University of Ireland Galway, Galway, Ireland.}
\date{Received 20 May 2005 / Accepted 21 July 2005}

\abstract{Optical CCD imaging and spectroscopic observations of three
supernova remnants are presented. Optical emission from G 54.4$-$0.3
and G 59.8$+$1.2 is detected for the first time, while the first flux
calibrated CCD images of the supernova remnant G 126.2$+$1.6 were
performed in the optical emission lines of \hnii, \oiii\ and \sii. A
mixture of filamentary and diffuse structures is observed in G 54.4$-$0.3
and G 59.8$+$1.2, mainly in \hnii, while the deep optical images of G
126.2$+$1.6 reveal several new filamentary and diffuse structures
inside the extent of the remnant as defined by its known radio
emission. In all cases, the radio emission is found to be well
correlated with the optical filaments. \oiii\ emission was not
detected at G 54.4$-$0.3 and G 59.8$+$1.2 while in G 126.2$+$1.6,
significant morphological differences between the low and medium
ionization images are present suggesting incomplete shock
structures. Deep long--slit spectra were taken at different positions
of the remnants. Both the flux calibrated images and the long--slit
spectra clearly show that the emission originates from shock--heated
gas, while some spectra of G 126.2$+$1.6 are characterized by large
\oiii/\hbeta\ ratios. This remnant's \oiii\ flux suggests shock velocities 
into the interstellar ``clouds'' between 100 and 120 \vel, while the
\oiii\ absence in the other two remnants indicates slower shock
velocities. For all remnants, the \siirat\ ratio indicates electron
densities below 600 \dens with particularly low densities for G
54.4$-$0.3 (below 50 \dens). Finally, the \ha\ emission has been
measured to be between 3.0 to 15.2 $\times$ \flux, 3.2 $\times$ \flux\
and between 6.5 to 16.8 $\times$ \flux\ for G 54.4$-$0.3, G 59.8$+$1.2
and G 126.2$+$1.6, respectively.
 }

\maketitle

\section{Introduction}
The majority of the galactic supernova remnants (SNRs) have been
identified by their synchrotron emission while some of the remnants
have been discovered in the soft X--rays and in the optical
band. Optical emission line observations of SNRs show that the
\sulfur/\ha\ ratio is typically higher than $\sim$0.4, while in
photoionized nebulae the ratio drops below 0.3--0.4 (Smith \et\
\cite{smi93}). Additional line ratios are used in doubtful cases.

The galactic supernova remnant G 126.2$+$1.6 was discovered in a 1420
MHz radio continuum survey by Reich \et\ (\cite{rei79}), where it
appears as a large but not well defined shell ($\sim$68\arcmin\ in
diameter) of low surface brightness. Radio observations were also
performed at 408, 865, 1410, 2700 and 4850 MHz by Reich \et\
(\cite{rei03}), Joncas \et\ (\cite{jon89}) and F\"{u}rst \et\
(\cite{fur84}) and a spectral index of 0.5--0.6 was established.
Optical filaments which coincide with the brightest radio emission on
the west part of the remnant have been detected by Blair \et\
(\cite{bla80}), Rosado (\cite{ros82}) and Fesen \et\ (\cite{fes83}),
while Xilouris \et\ (\cite{xil93}) presented a wide field \hnii\ image
(26\arcmin$\times$40\arcmin). In particular, imaging and spectroscopy
of one of the filaments were performed by Blair \et\ (\cite{bla80})
suggesting a large \oiii/\hbeta\ ratio, while Rosado (\cite{ros82})
was found relatively bright filamentary emission in \ha\ and \sii,
extending further to the north and south of the already known
filaments. \oiii\ images of the west part of the remnant were also
obtained by Fesen \et\ (\cite{fes83}). These images show faint
emission which appear correlated with the radio emission along the
western rim. G 126.2$+$1.6 is classified as an old remnant which still
shows strong \oiii\ emission (Raymond
\cite{ray84}).

The extended shell--type remnant G 54.4$-$0.3 was first identified by
Holden \& Caswell (\cite{hol69}) during their 178 MHz radio
survey. Several radio observations have been performed (Junkes \et\
\cite{jun92a} and references therein) showning its non-thermal nature,
a spectral index of $\sim $0.4 and a nearly circular shell (with a gap
near the north--east boundary). A distinct radio shell of
$\sim$40\arcmin diameter with a gap in its eastern part is seen in the
radio maps of Junkes \et\ (\cite{jun92a}). They also performed CO
observations and suggested a lower distance of $\sim$3 kpc to G
54.4$-$0.3. IRAS and high resolution CO observations of the
surrounding region (Junkes \et\ \cite{jun92b}) show an
OB--association, a complex of \HII\ regions and a CO--shell, all at a
distance of $\sim$3 kpc. They concluded that G 54.4$-$0.3 is associated
with these objects and is part of an extended complex of young
population I objects, where its progenitor star possibly was
born. X--ray emission from G 54.4$-$0.3 and the surrounding region was
detected by ROSAT (Junkes \cite{jun96}) which is nicely
anti--correlated with the cold molecular gas. A hydrogen column
density of 10$^{22}$ cm$^{-2}$~and a plasma temperature of $\sim
2\times 10^{7}$ K were estimated from the thermal X--ray emission.

G 59.8$+$1.2 was first detected by Reich \et\ (\cite{rei88}) in the
Effelsberg 2.7--GHz survey, while its radio image was published by
Reich \et\ (\cite{rei90}). It is classified as a supernova remnant with an
incomplete radio shell, an angular size of $\sim$20\arcmin
$\times$16\arcmin, and a spectral index of $\sim$0.5 (Green
2004). Radio surveys of the surrounding region, do not reveal any
pulsar to be associated with G 59.8$+$1.2. Neither of the last two
remnants has been detected optically in the past.

In this paper, we report the discovery of new faint optical
filamentary and diffuse emission from G 54.4$-$0.3, G 59.8$+$1.2 and G
126.2$+$1.6. We present \hnii\ and in some cases \sii~and \oiii~images
which reveal filamentary and diffuse structures well correlated with
the radio emission. Spectrophotometric observations of the brightest
filaments were also performed. In Sect. 2, we present information
concerning the observations and data reduction, while the results of
the imaging and spectral observations are given in Sect. 3, 4 and 5
for G 126.2$+$1.6, G 59.8$+$1.2 and G 54.4$-$0.3, respectively. In the
last section (Sect. 6) we discuss the physical properties of the
supernova remnants.

\section{Observations}
\subsection{Imaging}

The observations were performed with the 0.3 m Schmidt-Cassegrain
(f/3.2) telescope at Skinakas Observatory in Crete, Greece in July 7,
August 27 \& 28, September 23 \& 27, 2002 and June 25 \& 26, 2003. The
1024$\times $1024 (19 $\mu$m pixel) Thomson CCD camera was used
resulting in a scale of 4\arcsec\ pixel$^{-1}$~and a field of view of
70\arcmin~$\times$~70\arcmin.

A series of exposures in \hnii, \oiii\ and \sii\ each of 2400 s were
taken during the observations, resulting in different total exposure
times depending on the object. The details of all imaging observations
are given in Table~\ref{table1}. The final images in each filter are
the average of the individual frames.

The image reduction was carried out using the IRAF and MIDAS
packages. The astrometric solution for each field was calculated using
reference stars from the Hubble Space Telescope (HST) Guide Star
Catalogue (Lasker \et\ \cite{las99}). The spectrophotometric standard
stars HR5501, HR7596, HR7950, and HR8634 (Hamuy \et\ \cite{ham92})
were used for absolute flux calibration. All coordinates quoted in
this work refer to epoch 2000.

\subsection{Spectroscopy}
Low dispersion long--slit spectra were acquired with the 1.3 m
Ritchey-Cretien (f/7.7) telescope at Skinakas Observatory in July 8,
August 29, September 12, 2002 and June 29, July 3, 6 \& 7, 2003. The
1300 line mm$^{-1}$~grating was used in conjunction with a
2000$\times$800 SITe CCD (15 $\mu$m pixel) resulting in a scale of 1
\AA\ pixel$^{-1}$ and covers the range of 4750 \AA\ -- 6815 \AA. The
spectral resolution is $\sim$8 pixels and $\sim$11 pixels full width
at half maximum (fwhm) for the red and blue wavelengths,
respectively. The slit width is 7\farcs7 and in all cases was oriented
in the south--north direction; the slit length is 7\farcm9. Details
about the coordinates of the slit centers, the number of spectra and
their individual exposure times are given in Table~\ref{table1}. The
spectrophotometric standard stars HR5501, HR7596, HR9087, HR718, and
HR7950 were observed to calibrate the spectra.
\section{The supernova remnant G 126.2$+$1.6}
\subsection{The \hnii\ and \sii\ emission line images}
\begin{figure}
\resizebox{\hsize}{!}{\includegraphics{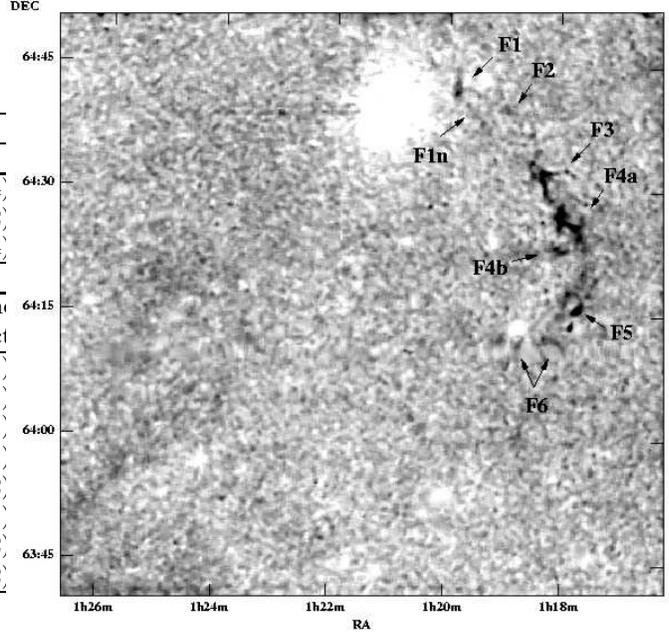}}
\caption{The field of G 126.2$+$1.6 in the \hnii\ filter. Labels
F1--F6 define the areas discussed in the text in more detail. Note
that F1n defines the area of the new \oiii\ filament which can be seen
clearly in Fig. \ref{figG126b}. The image has been smoothed to
suppress the residuals from the imperfect continuum
subtraction. Shadings run linearly from 0 to 100 $\times$ \flux.}
\label{figG126a} \end{figure}

Fig. \ref{figG126a} shows the \hnii\ image, where new faint emission
including the known filamentary structures, can be seen. In
particular, the image shows several thin and curved filaments all
present in the west, south--west and north--west areas of the remnant,
while no emission is detected in the east. In Table \ref{fluxes}, we
list typical fluxes measured in several locations within the field of
G 126.2$+$1.6. The morphology seen in the \sii\ filter is similar to
that in the \hnii\ and is not shown here. Both images being flux
calibrated provide a first indication of the nature of the observed
emission. An examination of the diagnostic ratio
\sii/\ha\ shows that the emission from the brightest part of the
remnant originates from shock--heated gas since we estimate ratios
\sii/\ha\ of 0.8--1.2, which are in agreement with our spectra
measurements (Sect. 3.3). Each filter contribution was estimated using
the methodology of Mavromatakis \et\ (\cite{mav02b}).
\par
Morphologically, starting from the north area, a bright filament
2\arcmin\ long (named F1 in Fig. \ref{figG126a}) is present which lies
a few arcminutes to the west of the variable star HD 8003 with its
center approximately at $\alpha \simeq$ 1\h19\m51\s\ and $\delta
\simeq$ 64\degr42\arcmin13\arcsec. South--west of this filament
appears a much fainter one (F2) at $\alpha \simeq$ 1\h18\m58\s,
$\delta \simeq$ 64\degr40\arcmin00\arcsec\ which is up to 1\arcmin\
long. Further to the south--west, a prominent new bright structure
which appears exactly to the north of the known optical filament in
the west (Rosado \cite{ros82}). This structure (named F3,
$\sim$7\arcmin\ long, $\sim$1\arcmin.2 wide) lies between $\alpha
\simeq$ 1\h18\m22\s, $\delta \simeq$ 64\degr33\arcmin10\arcsec\ and
$\alpha \simeq$ 1\h18\m08\s, $\delta \simeq$
64\degr28\arcmin41\arcsec\ has the same curvature and separated by a
$\sim$2\arcmin\ gap with the known west filament. A much more detailed
image of the known optical emission can be seen in
Fig. \ref{figG126a}. The west structure consist of two main parts
centered at $\alpha \simeq$ 1\h17\m48\s, $\delta \simeq$
64\degr25\arcmin58\arcsec\ (F4a) and $\alpha \simeq$ 1\h18\m02\s,
$\delta \simeq$ 64\degr22\arcmin53\arcsec\ (F4b) and separated by a
few arcminutes. Diffuse emission is also present to the east and south
of these structures. Another $\sim$8\arcmin\ long filament (F5) which
appears to the south has also the same curvature with the west
filament and lies between $\alpha \simeq$ 1\h17\m37\s, $\delta
\simeq$ 64\degr16\arcmin55\arcsec\ and $\alpha \simeq$ 1\h17\m50\s,
$\delta \simeq$ 64\degr15\arcmin14\arcsec. Diffuse emission appears
along north and south of this filament. Finally, there are extremely
faint new filaments (Area F6, $\sim$2--3\arcmin) in the area between
$\alpha \simeq$ 1\h17\m58\s -- 1\h19\m06\s\ and $\delta \simeq$
64\degr08\arcmin\ -- 64\degr12\arcmin\ which also seem to have the
same curvature with the other filaments.
\subsection{The \oiii\ emission line image}
\begin{figure}
\resizebox{\hsize}{!}{\includegraphics{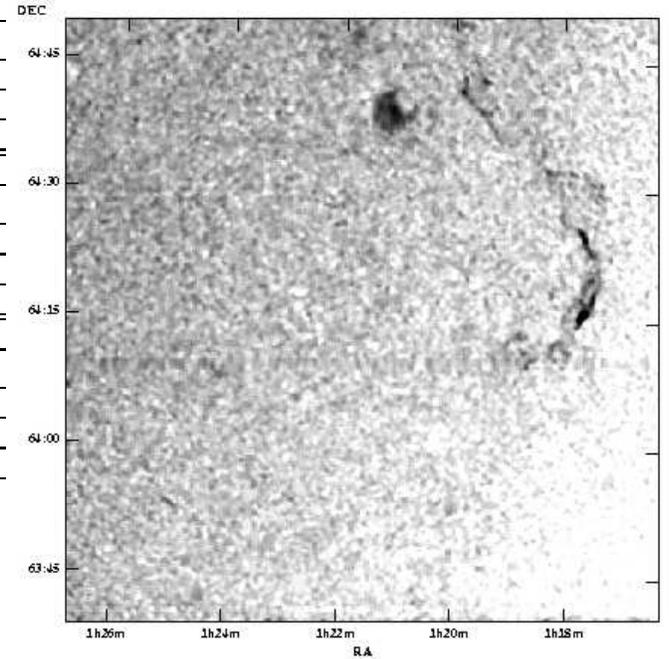}}
\caption{The field of G 126.2$+$1.6 in the \oiii\ filter. The image
has been smoothed to suppress the residuals from the imperfect
continuum subtraction. Shadings run linearly from 0 to 18 $\times$
\flux.}  
\label{figG126b} 
\end{figure}

\begin{figure}
\resizebox{\hsize}{!}{\includegraphics{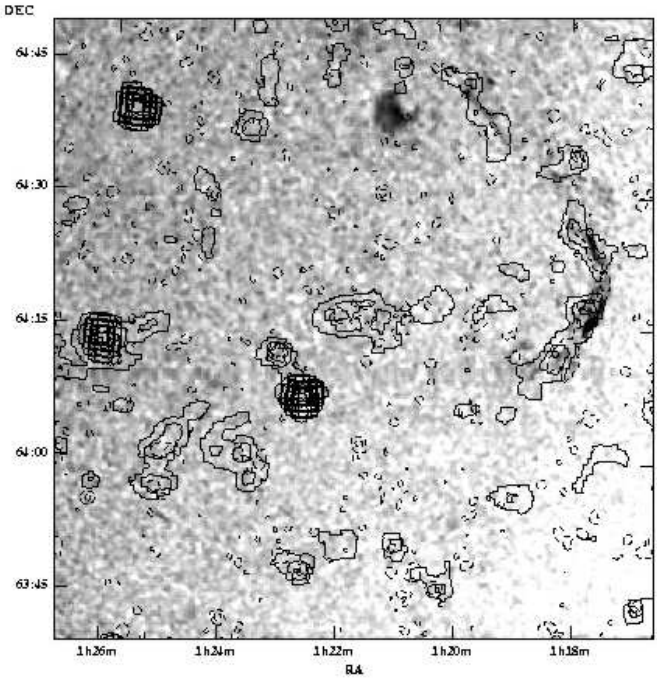}}
\caption{The correlation between the \oiii\ emission and the radio
emission at 1400 MHz (dashed line) and 4850 MHz (solid line) for G
126.2$+$1.6 is shown in this figure. The 1400 MHz and the 4850 MHz
radio contours scale linearly from 1.1$\times$10$^{-3}$~to
0.01 Jy/beam (resolution 45\arcsec$\times$45\arcsec) and
7$\times$10$^{-3}$~to 0.1 Jy/beam (7\arcmin$\times$7\arcmin),
respectively.}
\label{figG126c} 
\end{figure}

The detected \oiii\ emission (Fig. \ref{figG126b}) appears more
filamentary and less diffuse than in the \hnii\ image. In
Table~\ref{fluxes} typical \oiii\ fluxes are listed. Significant
differences between the \hnii\ and
\oiii\ images are present for most of the filaments. In particular, to
the north in contrast to the 2\arcmin\ filament (F1) found in \hnii,
there is a new faint thin filament (F1n -- 13\arcmin\ long) which lies
between $\alpha \simeq$ 1\h19\m56\s, $\delta \simeq$
64\degr42\arcmin15\arcsec\ and $\alpha \simeq$ 1\h19\m11\s, $\delta
\simeq$ 64\degr35\arcmin58\arcsec\ and follows the same curvature with
the west filament. This new long filament is also very well correlated
with the 4850 MHz radio map (Fig. \ref{figG126c}). On the other hand,
the new 7\arcmin\ bright \hnii\ filament (F3) does not have \oiii\
counterpart and at the same location only very faint diffuse emission
is found. The known west structure F4 (detected in \oiii\ by Fesen
\et\ \cite{fes83}) appears also bright but thinner in \oiii\ than
\hnii. The 8\arcmin\ filament F5 (detected in \oiii\ by Blair \et\
\cite{bla80}; Fesen \et\ \cite{fes83}) displays a different morphology
between the lower ionization lines of
\hnii\ and \sii\ and the medium ionization line of \oiii\ where it is
much brighter and better defined. Finally, A similar situation appears
both in \oiii\ and \hnii\ in the areas where weak and diffuse emission
is found.
\subsection{The optical spectra from G 126.2$+$1.6}
The deep low resolution spectra were taken on the relatively bright
optical filaments at two different locations (Table~\ref{table1}). In
Table~\ref{sfluxes}, we quote the relative line fluxes taken from the
above locations (designated Area I and II). In particular, in Area I,
we extracted two different apertures (Ia and Ib) along the slit that
are free of field stars and include sufficient line emission to allow
an accurate determination of the observed lines. The background
extraction apertures were selected towards the north and south ends of
each slit depending on the filament's position within the slit. The
measured line fluxes indicate emission from shock heated gas, since
\sii/\ha\ $\simeq$1.0. The spectra indicate significant attenuation of the
optical radiation but given the low counting statistics we cannot
attribute the extinction variations to intrinsic absorption of the
object. The signal to noise ratios do not include calibration errors,
which are less than 10 percent.
\par
The absolute \ha\ flux covers a range of values, from 6.5 to 16.8
$\times$ \flux. The \siirat\ ratio which was calculated between 1.1
and 1.4, indicates electron densities between 30 to 400 cm$^{-3}$~
(Osterbrock \cite{ost89}). However, taking into account the
statistical errors on the sulfur lines, we calculate that electron
densities up to 600 cm$^{-3}$ are allowed (Shaw \& Dufour
\cite{sha95}). Furthermore, as noted above, \hbeta\ emission was detected in
the spectra of Area Ib and II while in filament of Area Ia only an
upper limit is given. However, measurements for the \oiii/\hbeta\
ratio result in values larger than 30 (Area I) and less than 6 (Area
II). Theoretical models of Cox \& Raymond (\cite{cox85}) and Hartigan
\et\ (\cite{har87}) suggest that for shocks with complete
recombination zones this value is $\sim$6, while this limit is
exceeded in case of shock with incomplete recombination zones (Raymond
\et\ \cite{ray88}). Our measured values suggest, in Area I that
shocks with incomplete recombination zones are present, while in Area
II shocks with complete recombination zones. Therefore, according to
our measurements and the above theoretical models the estimated shock
velocities are $\sim$100 \vel\ and $\sim$120 \vel\ for Area II and I,
respectively.
\subsection{Observations at other wavelengths}
Several radio observations of G 126.2$+$1.6 have been performed in the
past (see Sect. 1 for detail). The optical filaments match very well the
radio emission of G 126.2$+$1.6 at 1410 MHz (Reich \et\
\cite{rei79}) and 4850 MHz (F\"{u}rst \et\ \cite{fur84}), suggesting
their association (Fig. \ref{figG126c}). Note that a first indication
of such correlation was made by Blair \et\ (\cite{bla80}) who drawn a
sketch locating their optical \oiii\ filaments with respect to the
radio contours at 1400 MHz. The area of G 126.2$+$1.6 was observed by
ROSAT during the All--Sky survey but no emission was detected. A small
part of the remnant was in the field of view of a ROSAT pointed
observation (RP400291N00). The image of this observation reveals
extended emission ($\sim$7\arcmin.5$\times$15\arcmin) centered at
1\h21\m52\s, 64\degr32\arcmin50\arcsec. The typical flux of this
structure is $\sim$1.1$\times$10$^{-3}$ cts s$^{-1}$ arcmin$^{-2}$~and
seems to be present both in the 0.5--1.0 keV and the 1.0--2.4 keV
bands. However, it may be difficult to correlate this emission to G
126.2$+$1.6 because it is located close to the edge of the field of
view of the PSPC ($\sim$51\arcmin\ from the center). The IRAS maps
have also been searched at 12, 25, 60 and 100 $\mu$m but no features
which might suggest an interaction of the supernova remnant with an IR
source were found.
\section{The supernova remnant G 59.8$+$1.2}
\subsection{The \hnii, \sii\ and \oiii\ emission line images}
Optical filamentary and diffuse emission is detected for a first time
for this remnant. The major characteristic revealed from the \hnii\
image (Fig. \ref{figG59a}) seems to be the low surface brightness of G
59.8$+$1.2. Weak, diffuse emission is present in the south,
south--east and central areas of the remnant, while no emission is
detected in the north. The most interesting region lies in the west,
where a complex filamentary structure exists (between \a\ $\simeq$
19\h38\m30.7\s, \dd\ $\simeq$ 24\degr10\arcmin03\arcsec\ and
\a\ $\simeq$ 19\h38\m08.2\s, \dd\ $\simeq$ 24\degr15\arcmin38\arcsec),
which is very well correlated with the radio emission. This bright
filament extends for $\sim$8\arcmin\ in the west, south--west. In
Table \ref{fluxes}, we present typical average fluxes measured in
several locations within the field of G 59.8$+$1.2 including an
unknown \HII\ region which is located south and south--east of the
remnant. In contrast to the previous results, diffuse emission was
mainly detected in the \sulfur\ emission line image, while no
significant emission was found in the image of the \oiii\ medium
ionization line. We did not detect \sii\ emission where most of the
\hnii\ emission was found but only where the bright west filament
appears. Therefore, the \sii\ and \oiii\ images are not shown
here. Table \ref{fluxes}~lists typical \sii\ and \oiii\
(3--$\sigma$~upper limits) fluxes measured in different parts of the
remnant. A deeper study of these images shows that the emission from
the brightest part of the remnant (west filament) originates from
shock heated gas since we estimate a ratio \sulfur/\ha\
$\sim$0.4--0.6, while a photoionization mechanism acts in the
south--east region (\sii/\ha\ $\sim$0.2--0.3).

Assuming that the east filaments belong to the remnant, their geometry
allows us to approximately define its diameter. In particular, two
possibilities will be examined. If the east border of the remnant is
defined by the outer filament at \a\ $\simeq$ 19\h39\m53.0\s, \dd\
$\simeq$ 24\degr19\arcmin51\arcsec, then a diameter of 22\arcmin
$\times$ 20\arcmin.5 can be derived with its center at \a\ $\simeq$
19\h39\m04\s, \dd\ $\simeq$ 24\degr17\arcmin15\arcsec, while if the
inner filament at \a\ $\simeq$ 19\h39\m31.0\s, \dd\ $\simeq$
24\degr15\arcmin35\arcsec\ defines the remnant's east border then its
diameter is 19\arcmin $\times$ 16\arcmin\ with the centre now to be at
\a\ $\simeq$ 19\h39\m00\s, \dd\ $\simeq$
24\degr14\arcmin18\arcsec. Note that the latter optical angular size
is in very good agreement with the value of $\sim$20\arcmin\ $\times$
16\arcmin\ given in Green's catalogue (Green 2004). However, the
larger angular diameter cannot be excluded since the infrared emission
(Sect. 4.3) is in agreement with the larger diameter value, X--ray
emission has not been detected so far and the radio shell is
incomplete.
\subsection{The spectra of G 59.8$+$1.2}
A low resolution spectrum was taken at a bright area (see Table 1 for
position details) among the diffuse emission seen in the \hnii\ image,
which coincides with the non-thermal radio emission. The selected
aperture and background are chosen using the same criteria as for G
126.2$+$1.6 (Sect. 3.3). The measured fluxes are given in Table 3. The
\sii/\ha\ ratio of 0.68 shows that the optical radiation originates
from shocked gas. \oiii\ emission was not detected while the very weak
\hbeta\ emission suggests significant interstellar attenuation of
the optical emission. Using the latter emission, the upper limit of
the interstellar extinction c(\hbeta) is 1.85. The absence of \oiii\
emission can be explained by slow shocks ($\leq$ 70 \vel) since the
presence of higher velocities shocks would produce detectable \oiii\
emission. Furthermore, the \hbeta\ emission upper limit and the
absence of \oiii, result in values of the \oiii/\hbeta\ ratio of less
than 1. This suggests that shocks with complete recombination zones
are likely to be present. The absolute \ha\ flux was measured to be
3.2 $\times$ \flux. The \siirat\ ratio which was calculated to 1.06,
indicates electron densities to 470 cm$^{-3}$ according to the task
``temden'' in the nebular package in IRAF (Shaw \& Dufour
\cite{sha95}). However, taking into account the statistical errors on
the sulfur lines, we calculate that electron densities as low as 380
cm$^{-3}$ and as high as 580 cm$^{-3}$ are allowed.
\subsection{Observations at other wavelengths}
\begin{figure}
\resizebox{\hsize}{!}{\includegraphics{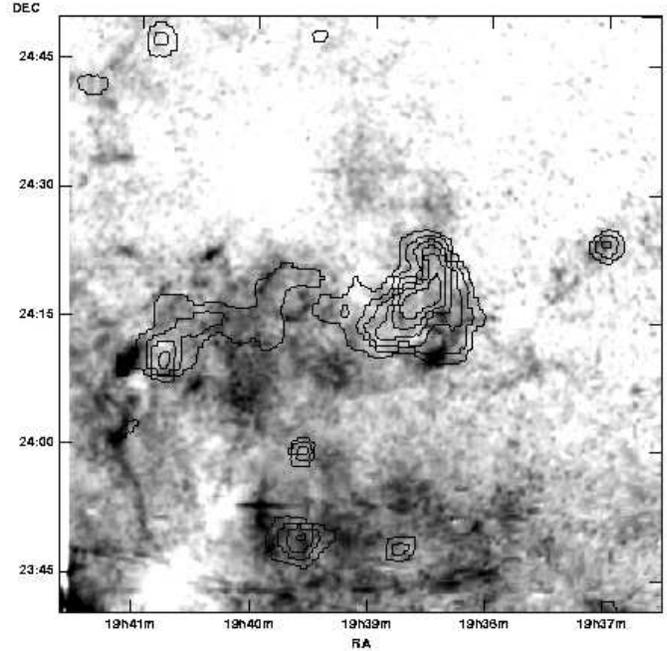}}
\caption{The correlation between the \hnii\ emission of G 59.8$+$1.2
and the radio emission at 4850 MHz (solid line) is shown in this
figure. The image has been smoothed to suppress the residuals from the
continuum subtraction. Shadings run
linearly from 0 to 120 $\times$ \flux. The 4850 MHz radio contours
scale linearly from 2.0$\times$10$^{-2}$ Jy/beam to 0.15 Jy/beam.} 
\label{figG59a} 
\end{figure}

\begin{figure}
\resizebox{\hsize}{!}{\includegraphics{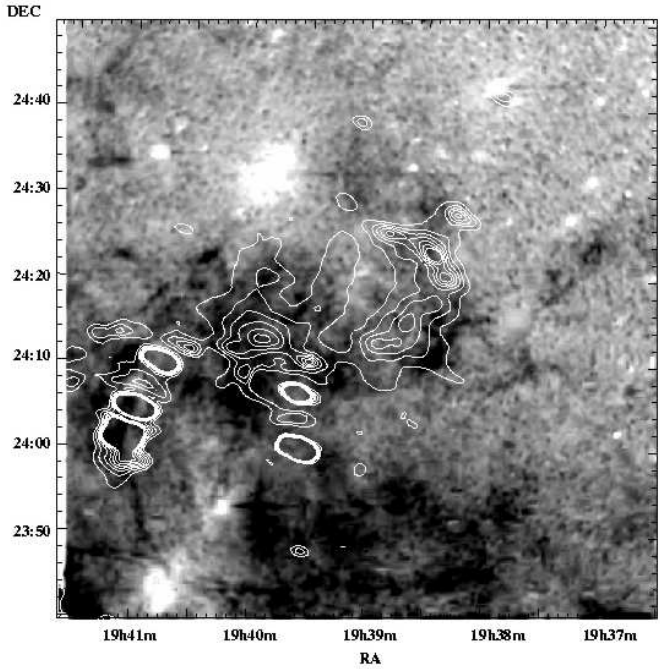}}
\caption{The correlation between the deep negative greyscale
representation of the \hnii\ emission and the IRAS 60 $\mu$m
(HiRes, white line) is shown in this figure. The \hnii\ image and the 60$\mu$m
contours scale linearly from 0 to 90 $\times$ \flux\ and from 8 to 40
MJy/sr, respectively.}
\label{figG59b} 
\end{figure}

The optical emission matches very well the radio emission of G
59.8$+$1.2 at 4850 MHz, suggesting their correlation
(Fig.~\ref{figG59a}). The observed filament is located close to the
outer edge of the radio contours but the low resolution of the radio
images does not allow us to determine the relative position of the
filament with respect to the shock front. In order to explore how the
infrared emission correlates with the optical emission, IRAS
resolution--enhanced (HiRes; Aumann \et\ \cite{aum90}) images at 60
$\mu$m of the same area were examined. Fig. \ref{figG59b} shows a deep
greyscale representation of the optical emission shown in
Fig.~\ref{figG59a} with overlapping contours of the infrared emission
(60 $\mu$m). IR emission correlated with G 59.8$+$1.2 is found towards
the south and west. The bright emission in the south--east coincides
with the optical emission therefore an association cannot be ruled
out, and especially in the case of a larger diameter remnant
(Sect. 4.1). Finally, X--ray emission was not detected in the ROSAT
All--sky survey.
\section{The supernova remnant G 54.4$-$0.3}
\subsection{The \hnii, \sii\ and \oiii\ emission line images}
\begin{figure}
\resizebox{\hsize}{!}{\includegraphics{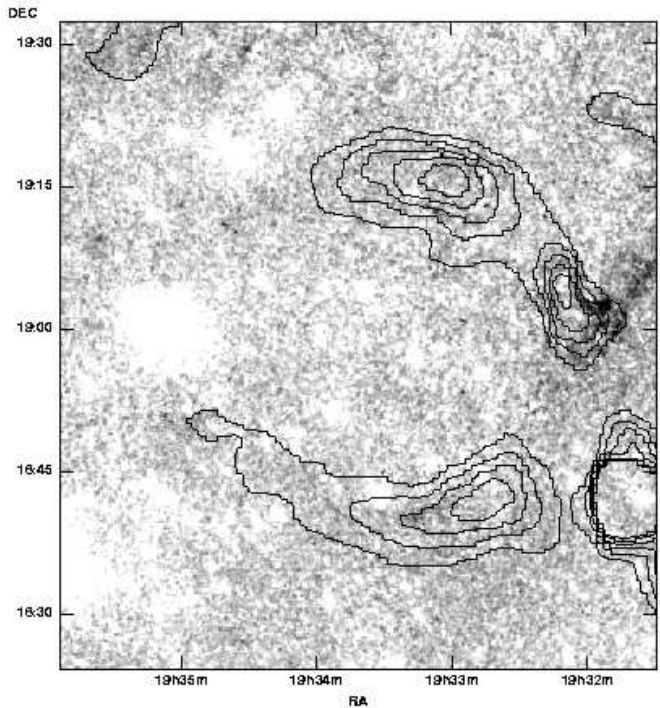}}
\caption{As Fig.~\ref{figG59a} but for G 54.4$-$0.3. Shadings run
linearly from 0 to 50 $\times$ \flux. The 4850 MHz radio contours
scale linearly from 3.5$\times$10$^{-2}$ Jy/beam to 0.25 Jy/beam.}
\label{figG54a}
\end{figure}

Optical emission from G 54.4$-$0.3 is detected for the first time as
for G 59.8$+$1.2. The \hnii\ image (Fig. \ref{figG54a}) shows
filamentary but also diffuse emission mainly in the west edge. The
morphology of the \sii\ image is generally similar to, though not as
bright as, that of the \hnii\ image and is not shown here, while no
significant \oiii\ emission was found. The flux calibrated images of
\hnii\ and \sii\ provide a first indication to the nature of the
observed emission (see Table 2). A study in different parts shows that
the emission from the west area originates from shock heated gas since
ratios \sii/\ha\ of $\sim$0.6--0.7 were found, while measurements of
very faint emission to the north and south of the remnant's edge show
a ratio \sii/\ha\ of $\sim$0.4--0.5 and $\sim$0.3--0.4,
respectively. It is possible that a photoionization mechanism acts in
these areas.
\subsection{The long--slit spectra from G 54.4$-$0.3}
The low resolution spectra cover the west (Area Ia-c), north (Area II)
and south (Area III) parts of the remnant (their exact positions are
given in Table 1). Spectra from Areas I \& II suggest that the
detected emission originates from shock heated gas (\sii/\ha\
$\sim$ 0.6), while no significant emission was detected in Area
III.  In general, the spectra of G 54.4$-$0.3 show similar results
with that of G 59.8$-$1.2 (Sect. 4.2) concerning the appearance of the
\hbeta\ and \oiii\ emission lines, hence the presence of slow shocks,
strong interstellar extinction and complete recombination zones can be
attributed. The Absolute \ha\ flux covers a range of values from 3.0
to 15.2 $\times$ \flux, while the \siirat\ ratio of 1.03 (Area II) and
1.43 (mean of Area Ia--c) indicate low electron densities (up to
$\approx$ 500 cm$^{-3}$).
\subsection{Observations at other wavelengths}
\begin{figure}
\resizebox{\hsize}{!}{\includegraphics{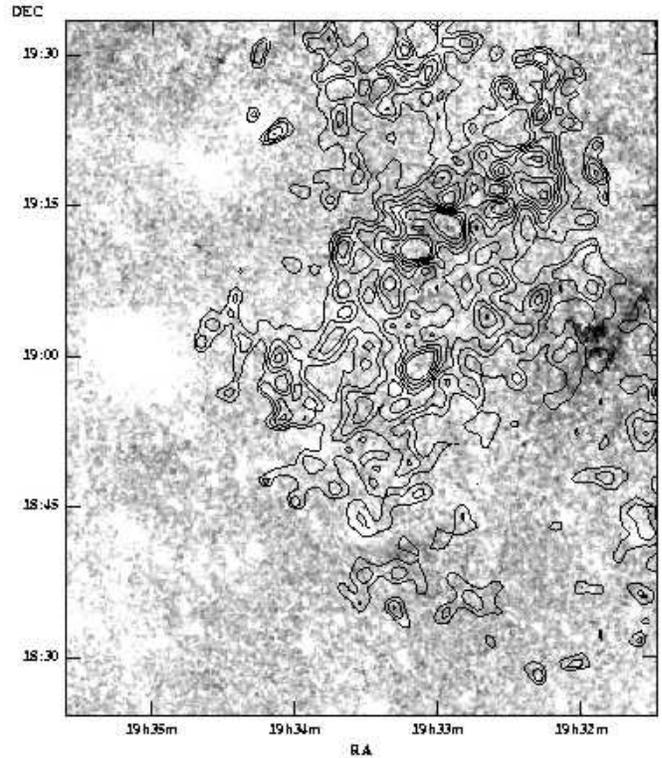}}
\caption{As Fig.~\ref{figG54a} but with the X--Ray ASCA at 0.5--10 keV
(contour) emission of G 54.4$-$0.3. The X--Ray contours scale linearly
from 2.1$\times$10$^{-4}$ Jy/beam to 3.5$\times$10$^{-4}$ Jy/beam.}
\label{figG54b}
\end{figure}

The newly discovered optical emission appears correlated with the
radio map of G 54.4$-$0.3 at 4850 MHz (Fig. \ref{figG54a}). The IRAS
(HiRes) images at 60 $\mu$m were also searched for emission features
that could be correlated with the optical ones. The detected IR
emission may not actually be in the area of the remnant but could be
part of a complex region (OB--association, \HII\ regions, CO--shell;
Junkes \et\ \cite{jun92b}). Archival ASCA data (Gotthelf \cite{got01})
have been used in order to investigate further the remnant's physical
properties. G 54.4$-$0.3 was observed by the ASCA satellite on October
14--16, 1994 with the GIS3 detector covering the energy band of
0.5--10 keV. Three sets of ASCA data pointed toward the centre, north
and southwest portion of the remnant were jointed to create the final
mosaic (Fig. \ref{figG54b}). Diffuse X--ray emission is present in the
area of this remnant correlates well with the optical emission found
to the west.
\section{Discussion}
The supernova remnants G 59.8$+$1.2 and G 54.4$-$0.4 are among the
least observed remnants in optical wavelengths while the G 126.2$+$1.6
new optical data improve the knowledge we had so far concerning its
physical properties.

\subsection{G 126.2$+$1.6}
This remnant shows up as an incomplete circular shell in optical and
radio wavelengths. Faint diffuse X--ray emission is also detected in
the area but its relation to G 126.2$+$1.6 is not clear. Our images
show the known emission and reveal more new filamentary and diffuse
structures. The new filament detected in the north, north--west is
very well correlated with the radio emission.  The calibrated images
as well as the long--slit spectra suggest that the detected emission
results from shock--heated gas. This remnant may belong to the group
of remnants which display partially strong \oiii\ emission like
e.g. CTB1 (Fesen \et\ \cite{fes97}), G 114.3$+$0.3 (Mavromatakis \et\
\cite{mav02a}), G 17.4$-$2.3 (Boumis \et\ \cite{bou02}). The morphological
differences between the low and medium ionization lines (Sects. 3.1,
3.2) suggest the presence of significant inhomogeneities and density
variations in the preshock medium (Hester
\cite{hes87}). Despite the low emission detected at \hbeta, an
estimation of the \oiii/\hbeta\ ratio suggest values greater than 30
(Area I) and less than 6 (Area II). Large
\oiii/\hbeta\ ratios are not unusual, since they have been measured by 
Fesen \et\ (\cite{fes83}) and Mavromatakis \et\ (\cite{mav02b}) in G
65.3$+$5.7 indicating shocks with incomplete recombination zones to be
present. The observed variation in the absolute line fluxes could be
due to variations of the interstellar cloud densities, shock
velocities or intrinsic absorption.
\par
A hydrogen column density N$_{\rm H}~{\rm
between}~7.8~{\rm and}~9.4 \times 10^{21}$~cm$^{-2}$ is given by
Dickey \& Lockman (\cite{dic90}) in the direction of the optical
filaments. Using the statistical relation of Predehl \& Schmitt
(\cite{pre95}), we obtain an N$_{{\rm H}}$~of $5.0 \times 10^{21}~{\rm
cm}^{-2}$~and $8.3 \times 10^{21}~{\rm cm}^{-2}$~for the minimum and
maximum c values calculated from our spectra (Table~\ref{sfluxes}),
respectively. Furthermore, using the E$_{B-V}$~value adopted from the
SFD code (Schlegel \et\ \cite{sch98}) an N$_{{\rm H}}$~of $7.9
\times 10^{21}~{\rm cm}^{-2}$ was calculated. Both values (our spectra
and SFD) are consistent with the estimated galactic N$_{{\rm
H}}$~considering the uncertainties involved.
\par
Joncas et al. (1989) argue that the distance to G 126.2$+$1.6 is
most likely to lie in the range 2 to 5 kpc. Blair et al (1980)
suggested that a missing hole of H{\sc i} lying at $2.4\pm 0.4~{\rm
kpc}$ could correspond to the ionized volume of G
126.2$+$1.6. Following Dopita (1979), a lower limit to the preshock
density (magnetic fields not included) can be estimated as $n_0 >
13.3~{\rm cm^{-3}}$ and the energy of the SN explosion is then
$E_{51}>0.26~D_{\rm kpc}^3\sim7 \times 10^{51}~{\rm ergs}$. Inputting
all of these results into a standard Sedov model (see, e.g., the
Appendix of Claas et al. 1989) yields the shock speed of $\sim
100~{\rm km~s^{-1}}$ and postshock temperature of $~5\times 10^5~{\rm
K}$.

At this temperature, the cooling time is short so the shock should be
approximately isothermal; the density constrast between the postshock
($n_1=200~{\rm cm^{-3}}$) and preshock gas ($n_0=13~{\rm cm^{-3}}$) in
an isothermal shock is related to the Mach number by $M2=n_1/n_0=15$
whereas the relation for an adiabatic shock is simply $n_1/n_0=4$
(Dyson \& Williams 1997). If the shock were non--magnetized and
fully isothermal then $M2$ would be at least 100 for a $100~{\rm
km~s^{-1}}$ shock. The value of $M2\sim 15$ could imply that the shock
is not quite isothermal which would be somewhat surprising given that
this is a large evolved remnant. If a strong ordered magnetic field
was present in the preshock gas then the swept up magnetic field would
limit the total compression in the postshock gas (Raymond \et\
\cite{ray88}) in those directions where the field lines were
originally parallel to the shock front.

\subsection{G 59.8$+$1.2}
The newly discovered filamentary and diffuse structures towards this
remnant show up as incomplete circular structures in the radio,
optical and infrared emission. X--ray emission has not been detected
so far. They provide a first evidence on the nature of the emission in
the area of the remmant suggesting that this emission originates from
shock--heated gas resulting from the interaction of the primary blast
wave with interstellar ``clouds''. A new filamentary structure,
unknown up to now, has been detected in the low ionization image of
\hnii\ to the south--east of G 59.8$+$1.2. However, the \sii/\ha\
ratio suggest that it is an \HII\ structure without any relation to
this remnant (Table 2). It is also situated well outside the radio
contours of the remnant. The spectrum of G 59.8$+$1.2 indicates that
the observed shock structures are complete since the
\oxygen/\hbeta\ $<$ 6 (Raymond \et\ \cite{ray88}). The
absence of soft X--ray emission may indicate a low shock temperature
and/or a low density of the local interstellar medium. The failure to
detect X--ray emission and the high interstellar extinction prevents
the determination of the local ISM density and explosion energy. The
\oxygen\ flux production depends mainly upon the shock velocity and
the ionization state of the preshocked gas. The absence of
\oxygen\ emission cannot help to determine whether slow shocks travel
into ionized gas or whether faster shocks travel into neutral gas (Cox \&
Raymond \cite{cox85}) but we can exclude moderate or fast shocks
overtaking ionized gas.
\par
Since there are no reliable \hbeta\ measurements for this remnant, we
adopt the statistical relation of Predehl \& Schmitt (\cite{pre95}),
using an E$_{B-V}$ calculated from the SFD code. Hence, we obtain an
N$_{{\rm H}}$~between 1.2 and 2.8 $\times 10^{22}~{\rm cm}^{-2}$.
Furthermore, according to the {\it FTOOLS} command ``nh'' (Dickey \&
Lockman \cite{dic90}), a total galactic N$_{{\rm H}}$ value of 1.3
$\times 10^{22}~{\rm cm}^{-2}$ is measured. However, it is worth
mentioning that using the upper limit of c(\hbeta), calculated from
our spectrum, we calculate an N$_{{\rm H}} >$ 2.5 $\times 10^{22}~{\rm
cm}^{-2}$ which is in agreement with the above calculations.
\par
Infrared emission (HiRes 60 $\mu$m) is found in the area of G
59.8$+$1.2. In Fig. \ref{figG59b}, the optical emission is shown with
contours of the IRAS emission where it can be seen that dust is
clearly associated with the remnant (Fig. \ref{figG59a}). A
correlation between the non--thermal radio emission and the infrared
is expected (Saken \et\ \cite{sak92}). It is also known that the
major, if not the primary, contribution to the infrared emission is
from shock--heated interstellar dust swept--up by the supernova
shell. Infrared emission is more sensitive to highly evolved remnants,
because it can detect shock--heated and radiatively--heated dust in
the shells of older remnants. The latter, usually have weak radio,
optical, UV and X--ray emission due to low shock velocities but still
relatively bright infrared emission (Shull
\et\ \cite{shu89}). If this case applies to G 59.8$+$1.2, then it could
be suggested that it is an evolved remnant.
\subsection{G 54.4$-$0.4}
Like G 59.8$+$1.2, optical emission is detected for the first time for
this remnant. The optical emission cannot define its
shape whilst an almost circular structure can be adopted from its
non--thermal radio emission. Both optical and radio emission do not
uniquely identify the nature of the detected emission, however, the
optical observations suggest the existence of shock--heated
structures. The \sii/\ha\ ratio of $\sim$ 0.6 could be explained by a lower
ionization state of the preshocked gas and/or a stronger magnetic
field (Raymond \cite{ray79}; Cox \& Raymond \cite{cox85}). Almost all
ratios of \siirat\ are close to the upper end of the allowable range
of values suggesting very low electron densities ($< 50$ \dens). This
fact along with the absence of the \oiii\ emission, the
shock modelling of Hartigan \et\ (\cite{har87}) and Raymond
\et\ (\cite{ray88}) point to shock velocities of $\leq$ 90 \vel 
(probably around 70--80 \vel) and low preshock clound densities, of the
order of a few atoms per cm$^{3}$.
\par
The interstellar extinction is not accurately determined due to the
low significance of the \hbeta\ flux. However, the lower limits on
c(\hbeta), which are derived from our spectra, suggest an area of high
interstellar extinction. A value for N$_{{\rm H}}$~between 2.9 and 4.0
$\times 10^{22}~{\rm cm}^{-2}$ is calculated from our data, while the
SFD code and the {\it FTOOLS} command ``nh'', result in values of
1.3--3.3 $\times 10^{22}~{\rm cm}^{-2}$~and 1.5 $\times 10^{22}~{\rm
cm}^{-2}$, respectively. Junkes (\cite{jun96}) using the soft X--ray
emission towards G 54.4$-$0.3, calculated a hydrogen column density of
10$^{22}$ cm$^{-2}$. Therefore, a value between 1.0 and 4.0 $\times
10^{22}~{\rm cm}^{-2}$~could be suggested for this remnant. The X--ray
emission from G 54.4$-$0.3 and the surrounding region shows a nice
anti--correlation with the cold molecular gas (CO; Junkes
\cite{jun96}). Diffuse X--ray emission is also detected by ASCA in the
area of the remnant (Fig. \ref{figG54b}) which correlates well with
the optical emission found in the west.
\par
IRAS and high resolution CO observations of the surrounding region
(Junkes \et\ \cite{jun92b}) show an OB--association, a complex of
\HII\ regions and a CO-shell, all at the same distance and an
association was made with G 54.4$-$0.3 suggesting that it is part of an
extended complex of young population I objects, where its progenitor
star possibly was born. According to Chu (\cite{chu97}), when a
supernova event occurs in a hot, low--density medium as is probably
the case, X--ray diffuse emission is expected to be present. This is
the case for supernovae with massive stars progenitors which are
usually formed in groups, such as OB associations. Therefore, the
current X--ray, optical, radio and CO information give rise to the
possibility that this remnant lies within an OB association.

\section{Conclusions}
Three supernova remnants were observed in major optical emission
lines. Emission was discovered for the first time in G 59.8$+$1.2 and
G 54.4$-$0.3, while previously unknown filamentary and diffuse
structures have been discovered in G 126.2$+$1.6. In all cases, the
optical emission is well correlated with the non-thermal radio
emission while infrared emission which might be related to the
remnants was found only in G 59.8$+$1.2. Diffuse X--ray emission
(ASCA) is also detected in the area of G 54.4$-$0.3 which probably
associates with the optical line emission, while in the case of G
126.2$+$1.6, X--Ray (ROSAT) emission was also found in the area but in
order to reach definite conclusions about its association to the
remnant, new X--ray observations are needed. No X--ray emission was
detected in the area of G 59.8$+$1.2. The images and the long--slit
spectra indicate that the emission arises from shock heated gas.
\begin{acknowledgements}
The authors would like to thank the referee (J. Raymond) for his
comments and suggestions. We would also like to thank Eric Gotthelf
who kindly provided us the ASCA image of G 54.4$-$0.3 in FITS
format. Skinakas Observatory is a collaborative project of the
University of Crete, the Foundation for Research and Technology-Hellas
and the Max-Planck-Institut f\"ur Extraterrestrische Physik. This
research has made use of data obtained through the High Energy
Astrophysics Science Archive Research Center Online Service, provided
by the NASA/Goddard Space Flight Center.
\end{acknowledgements}
%

\newpage

\begin{table}
\caption[]{Imaging and Spectral log.}
\label{table1}
\begin{flushleft}
\begin{tabular}{lccc}
\noalign{\smallskip}
\hline
\multicolumn{4}{c}{IMAGING} \\
\hline
Object & \hnii\ & \sii\ & \oiii\ \\
\hline
G 54.4$-$0.3 & 4800$^{a}$ (2)$^{b}$ & 12000 (5) & 9600 (4) \\
G 59.8$+$1.2 & 4800 (2) & 4800 (2) & 4800 (2) \\
G 126.2$+$1.6 & 4800 (2) & 9600 (4) & 9600 (4) \\
\hline
\multicolumn{4}{c}{SPECTROSCOPY} \\
\hline
Area & \multicolumn{2}{c}{Slit center} & Exp. time$^{\rm a}$ \\
 & R.A. & Dec. &  (No of spectra$^{c}$)\\
\hline
G 54.4$-$0.3 (Area Ia) & 19\h31\m54\s & 19\degr01\arcmin20\arcsec  & 3900 (1) \\
G 54.4$-$0.3 (Area Ib) & 19\h31\m53\s & 19\degr01\arcmin13\arcsec  & 3900 (1) \\
G 54.4$-$0.3 (Area Ic) & 19\h31\m53\s & 19\degr01\arcmin08\arcsec  & 3900 (1) \\
G 54.4$-$0.3 (Area II) & 19\h33\m06\s & 19\degr13\arcmin16\arcsec  & 3900 (1) \\
G 54.4$-$0.3 (Area III) & 19\h33\m14\s & 18\degr38\arcmin58\arcsec  & 7800 (2) \\
G 59.8$+$1.2 & 19\h38\m37\s & 24\degr07\arcmin45\arcsec  & 3900 (1) \\
G 126.2$+$1.6 (I) & 01\h17\m41\s & 64\degr15\arcmin58\arcsec  & 7200 (2) \\
G 126.2$+$1.6 (II) & 01\h17\m47\s & 64\degr26\arcmin32\arcsec  & 7800 (2) \\
\hline
\end{tabular}
\end{flushleft}
${\rm ^a}$ Total exposure time in sec.\\
${\rm ^b}$ Number of images obtained.\\
${\rm ^c}$ Number of spectra obtained.\\
\end{table}
\begin{table}
\caption[]{Typically measured fluxes over the brightest filaments.}
\label{fluxes}
\begin{flushleft}
\begin{tabular}{lllllll}
\hline
\multicolumn{7}{c}{G 54.4$-$0.3} \\
\hline
\noalign{\smallskip}
 & N  & W(1) & W(2) & S &  \\
\hline
\hnii\  & 9.5 & 29.6 & 20.0 & 11.0 & \\
\hline
\sii\   & 1.8 & 6.9 & 5.2 & 1.5 & \\
\hline
\oiii\ & \multicolumn{6}{c}{$<$7.3$^{\rm c}$}  \\
\hline
\hline
\multicolumn{7}{c}{G 59.8$+$1.2} \\
\hline
\noalign{\smallskip}
 & W & SW & SE & NE(I)$^{\rm a}$ & NE(O)$^{\rm a}$ & E$^{\rm b}$ \\
\hline
\hnii\  & 21.8 & 28.0 & 18.3 & 15.2 & 17.2 & 32.5\\
\hline
\sii\   & 3.8 & 5.9 & 2.4 & 5.2 & 11.8 & 5.8\\
\hline
\oiii\ & \multicolumn{6}{c}{$<$8.5$^{\rm c}$} \\
\hline
\hline
\multicolumn{7}{c}{G 126.2$+$1.6} \\
\hline
\noalign{\smallskip}
 & N  & NW & W & SW & S \\
\hline
\hnii\  & 14.2 & 17.7 & 26.3 & 29.0 & 5.8 \\
\hline
\sii\   & 4.8 & 7.0 & 8.2 & 8.5 & 2.6 \\
\hline
\oiii\  & 6.9 & 5.8 & 8.8 & 19.1 & 2.8 \\
\hline
\end{tabular}
\end{flushleft}
${\rm }$ Fluxes in units of \flux \\
${\rm}$ Median values over a 40\arcsec $\times$ 40\arcsec\ box. \\
${\rm ^a}$ Inner (I) and outer (O) filaments as defined in Sect. 4.1. \\
${\rm ^b}$ The unknown east bright region outside the snr's borders.\\ 
${\rm ^c}$ 3$\sigma$~upper limit.\\
 \end{table}

\begin{table*}
\caption[]{Relative line fluxes.}
\label{sfluxes}         
\begin{tabular}{lllllllll}
\hline
 \noalign{\smallskip}
 & \multicolumn{4}{c}{G 54.4$-$0.3} & G 59.8$+$1.2 & \multicolumn{3}{c}{G 126.2$+$1.6} \\
 & Area Ia & Area Ib & Area Ic & Area II &   & Area Ia$^{\rm 1}$ & Area Ib$^{\rm 1}$ & Area II  \\
Line (\AA) & F$^{\rm 2,3}$ & F$^{\rm 2,3}$ & F$^{\rm 2,3}$ & F$^{\rm 2,3}$ & F$^{\rm 2,3}$ & F$^{\rm 2,3}$ & F$^{\rm 2,3}$ & F$^{\rm 2,3}$  \\
\hline
4861 \hbeta\ & $<$ 6 & $<$ 5 & $<$ 3 & $<$ 3 & $<$ 8 & $<$ 6 & 12 (3) & 6 (2)  \\
4959 \oxygen\ & $-$ & $-$ & $-$ & $-$ & $-$ & 87 (15) & 114 (16) & $<$ 6 \\
5007 \oxygen\ & $-$ & $-$ & $-$ & $-$ & $-$ & 286 (39) & 372 (35) & 26 (10) \\
6548 \nii\ & 9 (3) & 18 (10) & 12 (10) & 26 (6) & 18 (20) & 24 (6) & 23 (3) & 17 (12) \\
6563 \ha\  & 100 (29) & 100 (57) & 100 (64) & 100 (19)  & 100 (84) & 100 (117) & 100 (15) & 100 (70) \\
6584 \nii\ & 35 (10) & 53 (30) & 35 (27) & 73 (14) & 51 (46) & 114 (31) & 107 (21) & 61 (42) \\
6716 \sii\ & 34 (10) & 40 (27) & 35 (27) & 31 (6) & 35 (32) & 52 (12) & 56 (10) & 54 (37) \\
6731 \sii\ & 26 (8) & 26 (19) & 24 (19) & 30 (6) & 33 (31) & 46 (18) & 50 (14) & 39 (26) \\
\hline
Absolute \ha\ flux$^{\rm 4}$ & 15.2 & 6.7 & 10.1 & 3.0 & 3.2 & 6.7 & 6.5 & 16.8 \\
\ha /\hbeta\  & $>$ 17 & $>$ 20 & $>$ 30 & $>$ 28 & $>$ 12 & $>$ 8 & 8.3 & 16.7 \\
\sii/\ha\     & 0.60 (14) & 0.67 (28) & 0.60 (29) & 0.60 (8) & 0.68 (39) & 0.98 (22) & 1.06 (12) & 0.93 (44) \\
F(6716)/F(6731)	& 1.34 (7) & 1.54 (15) & 1.48 (15) & 1.03 (5) & 1.06 (22) & 1.13 (10) & 1.12 (8) & 1.37 (22) \\
\oxygen/\hbeta\ & $-$ & $-$ & $-$ & $-$ & $-$ & $>$ 30 & 32.4 & 5.3 \\
c(\hbeta)$^{\rm 5}$ & $>$ 2.20 & $>$ 2.43 & $>$ 3.06 & $>$ 3.06 & $>$ 1.85 & $>$ 1.85 & 1.34 (3) & 2.21 (2) \\
\hline 
\end{tabular}

$^{\rm 1}$ Areas Ia and Ib, having an offset of 8\arcsec\ and
9\arcsec\ south and north of the slit center, respectively.

${\rm ^2}$ Listed fluxes are a signal to noise weighted
average of two fluxes.

${\rm ^3}$ Fluxes normalized to F(H$\alpha$)=100 and are uncorrected
for interstellar extinction.

$^{\rm 4}$ In units of \flux.

$^{\rm 5}$ The logarithmic extinction is derived by c =
1/0.348$\times$log((\ha/\hbeta)$_{\rm obs}$/2.85).

Numbers in parentheses represent the signal to noise ratio of the
quoted fluxes.\\
\end{table*}
%
%

\end{document}